\documentclass[manuscript,screen]{acmart}

\AtBeginDocument{%
 \providecommand\BibTeX{{%
 \normalfont B\kern-0.5em{\scshape i\kern-0.25em b}\kern-0.8em\TeX}}}

\copyrightyear{2023}
\acmConference[arXiv]{arXiv}{2023}{}

\usepackage{multirow}
\usepackage{colortbl}

\newlength\savedwidth

\begin{document}

\title[Error-rate Prediction for Mouse-based Rectangular-target Pointing with no Knowledge of Movement Angles]{Error-rate Prediction for Mouse-based Rectangular-target Pointing with no Knowledge of Movement Angles}

\author{Shota Yamanaka}
\affiliation{%
 \institution{Yahoo Japan Corporation}
 \city{Chiyoda-ku}
 \state{Tokyo}
 \country{Japan}
}

\renewcommand{\shortauthors}{Yamanaka}

\begin{abstract}
In rectangular-target pointing, movement angles towards targets are known to affect error rates.
When designers determine target sizes, however, they would not know the frequencies of cursor-approaching directions for each target.
Thus, assuming that there are unbiasedly various angles, we derived models to predict error rates depending only on the target width and height.
We conducted two crowdsourced experiments: a cyclic pointing task with a predefined movement angle and a multi-directional pointing task.
The shuffle-split cross-validation with 60\% training data showed $R^2 > 0.81$, $\mathit{MAE} < 1.3\%$, and $\mathit{RMSE} < 2.1\%$, suggesting good prediction accuracy even for predicting untested target sizes when designers newly set UI elements.
\end{abstract}


\maketitle

\section{Introduction}
To quantitatively determine how well users operate interactive systems, modeling error rates has attracted the interest of HCI researchers.
Developing error-prediction models for target pointing is a hot topic as this task is one of the most frequently performed actions on PCs and touch devices \cite{Bi16,Huang19uist,Usuba21iss,Yamanaka21hcomp}.

We derived error-rate prediction models on mouse-based pointing for rectangular targets.
Compared with traditional 1D ribbon-shaped or 2D circular targets, rectangular targets are recognized as more meaningful shapes in actual user interfaces (UIs) \cite{Accot03,Hoffmann94height,Ko20FF2D,Ma21uist,Yamanaka21crowdWH,Yang10,Zhang12}.
A notable difference of our work from previous mouse-based studies is that we did not use the initial cursor position of each trial, which prevented us from using the movement angle towards the target in the model.
This condition corresponds to a realistic UI design work, that is, designers want users to click a target with certain accuracy (e.g., $<$3\%) but would not know the frequencies of cursor-approach angles.

Conditions under which the approach angle is known are special cases, such as (a) a webpage has already been released and users' cursor traces are recorded or (b) a UI element is located close to the screen edge; thus, the frequencies are biased.
In more general cases, however, without prior knowledge, designers have no choice but to assume that the cursor reaches the target in various angles with no bias.

Error-rate prediction models that ignore movement angles have been proposed for touch-based pointing, in which a new target appears on the screen that participants tap \cite{Bi16,Yamanaka20issFFF}.
It could be more probable that the finger waits near the screen center than near screen edges, so the finger may tend to reach, e.g., a target on the top edge of the screen with an upward movement.
This means that not using movement angles is equivalent to merging various movement angles when predicting error rates \cite{Yamanaka20issFFF}.

Similarly to these touch-based pointing studies, we examined how well our error-rate models perform after merging various movement angles that our experimental system actually recorded.
We conducted two experiments: one with four movement angles and one with a multi-directional task modified from the ISO 9241-9 standard \cite{iso2000}.
The cross-validation results indicate that, even when the training-data size decreased by 40\%, the best-fit model could predict the error rates with $R^2 > 0.81$, mean absolute error ($\mathit{MAE}) < 0.7\%$, and root mean square error ($\mathit{RMSE}) < 0.8\%$ in Experiment 1, and $R^2 > 0.86$, $\mathit{MAE} < 1.3\%$, and $\mathit{RMSE} < 2.1\%$ in Experiment 2.

\section{Related Work}
\subsection{Predicting Error Rates in Pointing Tasks}
Modeling movement times ($\mathit{MT}$s) in pointing tasks has been a popular topic in HCI, namely, studies on Fitts' law \cite{Fitts54} and its variations (e.g., \cite{Accot03,Zhang12}).
In traditional experiments with 1D ribbon-shaped or 2D circular targets, the target size is solely defined by its width $W$ \cite{MacKenzie92,Soukoreff04}.
Under this condition, click- or touch-points (i.e., endpoints) are assumed to be distributed normally over the target \cite{Bi16,Crossman56,MacKenzie92,Wobbrock08error}, but this assumption does not always hold \cite{Welford69age,Yamanaka20issFFF}.

\begin{figure}[t]
\centering
\includegraphics[width=0.78\columnwidth]{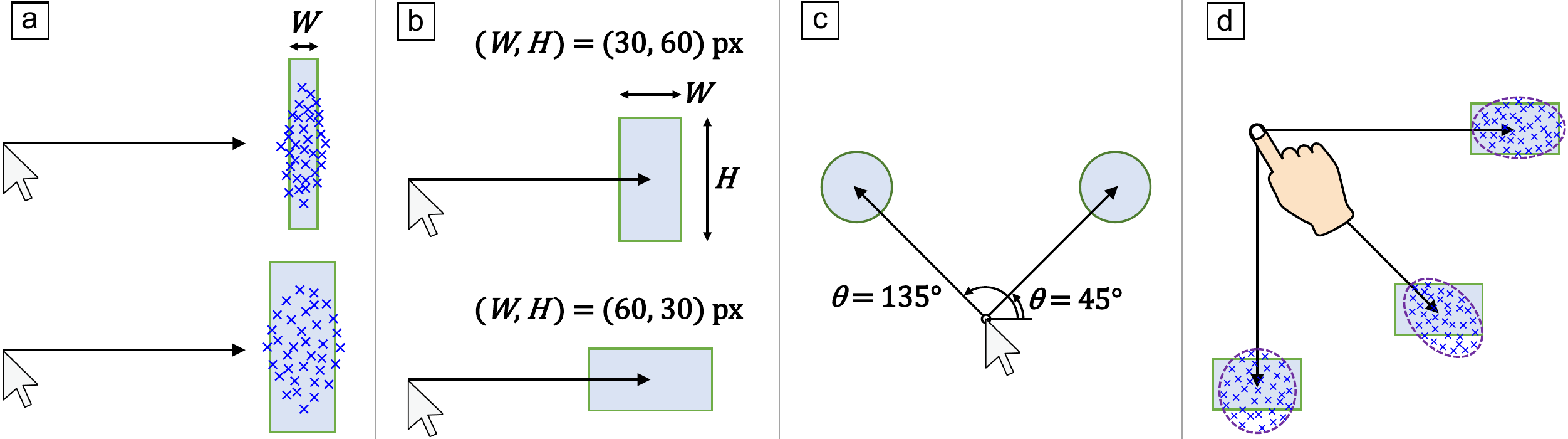}
\caption{Known relationships between task conditions and error rates or endpoint distributions. Endpoints are drawn as ``x''s. (a) The $\mathit{ER}$ increases as $W$ decreases. (b) Even when the movement angle and ratio of $W$ to $H$ are the same, the $\mathit{ER}$ is higher for the target with smaller movement-axis size, i.e., upper-row target. (c) Movement angles affect $\mathit{ER}$s. (d) Even for the same target sizes in terms of $W$ and $H$, the spread of endpoints tends to be lengthened along the movement axis.}
\label{fig:related}
\end{figure}

It is known that the error rate $\mathit{ER}$ tends to be higher as $W$ decreases (e.g., \cite{Fitts54,Chapuis11,Yamanaka21hcomp}; see Figure~\ref{fig:related}a).
However, this relationship between $\mathit{ER}$ and $W$ is not linear; thus, predicting the $\mathit{ER}$ requires a more probabilistic process.
The first step is to compute endpoint variability.
For normally distributed endpoints in 1D pointing on x-axis movements, the mean position is located close to the target center ($\mu_x = 0$), and the variability (standard deviation $\sigma_x$) increases with $W$.
It has been assumed that $\sigma_x$ is proportionally related to $W$ or with a small intercept:
\begin{equation}
\label{eqn:sigx_W_proportional}
\sigma_x = a + b\cdot{}W.
\end{equation}
Hereafter, italic lowercase letters $a$--$l$ refer to empirical constants.
This formulation has been used for predicting endpoint variability in mouse-based \cite{Yamanaka21hcomp} and virtual-reality pointing tasks \cite{YU19siggraph}.
As shown by this equation, the distance from the initial cursor (or finger) position to the target does not affect the endpoint distribution if users have sufficient time to aim for the target \cite{Bi16,Crossman56,Yamanaka20issFFF,Yamanaka21hcomp}.
We do not intend to run time-limited or ballistic (i.e., not using visual feedback) pointing tasks, where the target distance affects the distributions \cite{Beggs74,Grossman05prob,Lin15ballistic,Schmidt79,Wobbrock08error}.

The second step is to compute the $\mathit{ER}$ on the basis of $\sigma_x$.
By using the Gauss error function $\mathrm{erf}(\cdot{})$, the $\mathit{ER}$ is derived as
\begin{equation}
\label{eqn:ER_1D}
\mathit{ER} = 1 - \text{erf} \left( \frac{W}{2\sqrt{2} \sigma_x} \right).
\end{equation}
For a new target condition, we can accurately predict $\sigma_x$ by using Equation~\ref{eqn:sigx_W_proportional} then obtain the $\mathit{ER}$ on the basis of the new $W$ and predicted $\sigma_x$ by using Equation~\ref{eqn:ER_1D}.
For example, Yamanaka conducted a mouse-pointing task with $W$ ranging from 8 to 78 pixels and obtained $R^2 = 0.9966$ for Equation~\ref{eqn:sigx_W_proportional} and $R^2 = 0.9529$ for Equation~\ref{eqn:ER_1D} in leave-one-out cross-validation, which demonstrated good prediction accuracy for unknown target conditions \cite{Yamanaka21hcomp}.

\subsection{Endpoint Distribution and Error Rate in Rectangular Target Pointing}
Pointing to rectangular targets, which have another dimension (height $H$), has been studied to predict $\mathit{MT}$s \cite{Accot03,Bohan03bivariate,Crossman56,Hoffmann94height,Kvalseth77bivariate,Yamanaka21crowdWH}.
We define $W$ and $H$ as the target sizes on the x- and y-axes on the screen, respectively, as in these studies.

Endpoint distributions for rectangular targets have been studied in several contexts.
In a study on typing on a smartphone keyboard with $W = 6$ and $H = 9.375$ mm for each key, the tap points were reported as close to bivariate normal distributions \cite{Azenkot12}.
This was consistent with a study by Wang and Ren \cite{Wang09}.
In a game-like moving-target pointing task on a touchscreen, Huang et al. proposed the \textit{2D Ternary-Gaussian model} \cite{Huang19uist}.
This model includes more complex factors, such as finger-touch ambiguity and movement-speed effect, to model endpoint variability.

In Hoffmann and Sheikh's horizontal-movement experiment, they tested $W =$ 10, 20, and 40 mm, and $H =$ 1, 2, 5, 10, 20, 40, and 200 mm.
They reported that (a) $\sigma_x$ was significantly affected by $W$ but not by $H$, and the interaction between $W$ and $H$ was significant; and (b) the endpoint variability on the y-axis $\sigma_y$ was significantly affected by $H$ but not by $W$, and the interaction was not significant.
The results indicated that as either $W$ or $H$ increased, the $\sigma$ on that axis almost linearly increased.
The $R^2$ for ($W$, $\sigma_x$) was 0.91, and that for ($H$, $\sigma_y$) was 0.87.
After eliminating an extremely large $H$ of 200 mm, the $R^2$ for ($H$, $\sigma_y$) was 0.97.
Thus, for reasonable ranges where $W$ or $H$ constrains endpoints, we can assume that the endpoints spread more widely as the target size on that axis increases.

A similar conclusion was found in Yamanaka's dataset in a crowdsourcing experiment on a horizontal-movement task \cite{Yamanaka21crowdWH}.
He tested $W =$ 30, 40, 60, and 90 pixels, and $H =$ 10, 20, 30, 40, 60, 100, and 200 pixels.
The regressions of the 28 fitting points showed $R^2 > 0.96$ both for $(W, \sigma_x)$ and $(H, \sigma_y)$.
He also showed that, even when the movement angle and ratio of $W$ to $H$ are the same, the $\mathit{ER}$s are not consistent.
When $(W,H) = (30, 60)$ pixels, the $\mathit{ER}$ was 3.067\%, while that for $(60, 30)$ was 2.207\%, which showed a significant pairwise difference ($p < 0.001$); see Figure~\ref{fig:related}b.
Thus, the $\mathit{ER}$ would be higher when the movement-axis size is smaller.

The scope in both experiments by Hoffmann and Sheikh \cite{Hoffmann94height} and Yamanaka \cite{Yamanaka21crowdWH} was horizontal movements.
Thus, the validity of linear relationships for $(W, \sigma_x)$ and $(H, \sigma_y)$ should be empirically evaluated when we analyze movements with various angles.

\subsection{Effect of Movement Angle on Pointing Performance}
The movement angle towards the target is another objective in the pointing paradigm.
For modeling $\mathit{MT}$s, studies have shown that the times change depending on the angles, e.g., horizontal movements required a shorter time than vertical ones \cite{Appert08edge,Thompson04gainangle,Whisenand95angle,Yang10,Zhang12}.

In comparison, studies on $\mathit{ER}$s regarding the movement angle are rare.
One example is the work by Hertzum and Hornb{\ae}k who used circular targets and eight movement directions \cite{Hertzum10angle}.
They showed that the $\mathit{ER}$s for the top-right direction and top-left one were approximately 7.5 and 9\%, respectively, for right-handed mouse users (Figure~\ref{fig:related}c).
For touchpad usage, the downward movement showed the largest $\mathit{ER}$ of $\sim$13\%, and the smallest $\mathit{ER}$ of 10\% was observed for the rightward movement.
Thus, the $\mathit{ER}$ changes depending on the movement angle as well as the device.

Ma et al. proposed the \textit{Rotational Dual Gaussian Model} to predict touch-point distributions \cite{Ma21uist}.
They strictly controlled the movement angle in each trial and proposed a target-size adjustment method based on the movement angle, called \textit{projected target width and height}.
They showed that the endpoints were affected by $W$ and $H$, as well as the movement angle; the endpoint variability became larger on the movement axis than the perpendicular axis (Figure~\ref{fig:related}d).
They showed that their proposed model could estimate $\mathit{ER}$s more accurately than a model without using the angle.

Therefore, if researchers can distinguish each pointing trial's movement angle, the prediction accuracy of an $\mathit{ER}$ model would improve.
However, this is a kind of special case in which we can record, e.g., the cursor trajectory in a pointing-experiment system for research purposes, but the feasibility would be low for practitioners to use the model when they design UIs.
Hence, our research question is, if we cannot know the movement angles and we merge the data from various angles, can we accurately predict the $\mathit{ER}$s?

In summary, although the importance of $\mathit{MT}$ models for rectangular targets and developing $\mathit{ER}$ models have been recognized individually, $\mathit{ER}$ models for rectangular target pointing have rarely been studied.
Because Ma et al. have shown that integrating the movement angle significantly outperformed the angle-ignored model \cite{Ma21uist}, we focus only on the prediction accuracy when the movement angle is unknown, i.e., various angles appear unbiasedly and the data are analyzed after merging.

\section{Model Derivation}
As with the $\sigma_x$ regression, and as Hoffmann and Sheikh demonstrated \cite{Hoffmann94height}, we use the following equation to predict $\sigma_y$:
\begin{equation}
\label{eqn:sigy_H_proportional}
\sigma_y = c + d\cdot{}H.
\end{equation}
Because our assumption is that we cannot know the movement angle, $\sigma_x$ and $\sigma_y$ are independent from the movement axis, i.e., these are the endpoints' $\mathit{SD}$ values measured on the screen.
Equations~\ref{eqn:sigx_W_proportional} and \ref{eqn:sigy_H_proportional} are the simplest models to predict $\sigma_x$ and $\sigma_y$, but Hoffmann and Sheikh found that the interaction of $W$ and $H$ had a significant effect on $\sigma_x$.
As we later show, $H$ also had a main effect on $\sigma_x$, and $W$ had a main effect on $\sigma_y$.
Given that the interaction of $W$ and $H$ could be significant on $\sigma_x$ and $\sigma_y$, the possible more complicated formulations are
\begin{equation}
\label{eqn:sigx_WH}
\sigma_x = e + f\cdot{}W + g\cdot{}H + h\cdot{}\mathrm{interaction}(W,H)\ \ \ \mathrm{and}\ \ \ \sigma_y = i + j\cdot{}H + k\cdot{}W + l\cdot{}\mathrm{interaction}(W,H).
\end{equation}
The interaction terms will be determined from our experimental results.
If the effect of $W$ to increase $\sigma_x$ increases with $H$, the term should be $W \times H$.
In contrast, if the effect decreases as $H$ increases, it should be $W/H$.
Similarly, if the effect of $H$ to increase $\sigma_y$ decreases as $W$ increases, the interaction term is $H/W$.\footnote{Both multiplication and division can be used \cite{interactionterm18}, and previous studies used a similar empirical approach to determine the division form for the interaction or covariance term to model $\sigma$ \cite{Huang18error1D,Huang19uist,Zhou09temporal}. Another policy for variability fitting is to use the squares of variables then take a square-root, e.g., $\sigma_x = \sqrt{e + f\cdot{}W^2 + g\cdot{}H^2 + h\cdot{}\mathrm{cov}(W,H)}$. We tested both methods and found no notable difference; thus, we used the simpler one, as in previous studies \cite{Yamanaka21hcomp,YU19siggraph}.}

We assume that the endpoints follow a bivariate normal distribution, and the mean of click-points is located at the target center ($\mu_x = \mu_y = 0$).
Thus, the probability $P$ that a click position $(x, y)$ falls inside the target area $D$ is

\begin{equation}
\label{eqn:SR_original_corr}
 P(D) 
 = \iint_D \frac{1}{2 \pi\sigma_x \sigma_y \sqrt{1-\rho^2}} \text{exp}\left[-\frac{1}{2(1-\rho^2)} \left( \frac{x^2}{\sigma_x^2} + \frac{y^2}{\sigma_y^2} - \frac{2\rho x y}{\sigma_x \sigma_y} \right)\right] \mathit{dx}\mathit{dy},
\end{equation}
where $ D = \{ (x, y) | -\frac{W}{2} \leq x \leq \frac{W}{2}, -\frac{H}{2} \leq y \leq \frac{H}{2} \} $ and $\rho$ is the correlation for the endpoint distributions on the x- and y-axes.
Because we assume that $\rho$ is negligible ($\approx 0$) when merging various movement angles \cite{Bi16,Yamanaka20issFFF}, we have
\begin{equation}
\label{eqn:SR_original}
 \begin{split}
 P(D) 
 = \iint_D \frac{1}{2 \pi\sigma_x \sigma_y} \text{exp}\left(-\frac{x^2}{2 \sigma_x^2} - \frac{y^2}{2 \sigma_y^2} \right) \mathit{dx}\mathit{dy}
 = \iint_D \frac{1}{\sqrt{2 \pi} \sigma_x} \text{exp}\left(-\frac{x^2}{2 \sigma_x^2} \right) \frac{1}{\sqrt{2 \pi} \sigma_y} \text{exp}\left(-\frac{y^2}{2 \sigma_y^2} \right) \mathit{dx}\mathit{dy}.
 \end{split}
\end{equation}
Note that $\rho$ is concerning whether the endpoint distribution's ellipse is diagonal, while ``interaction$(W, H)$'' in Equation~\ref{eqn:sigx_WH} is concerning if the variability size increases with the interaction term.
Because the target area's x and y ranges are independent, the double integrals can be split, and the former and latter halves are written by Gauss error functions:
\begin{equation}
 \label{eqn:SR_simple}
 P(D) = \left[ \int_{- \frac{W}{2}}^\frac{W}{2} \frac{1}{\sqrt{2 \pi} \sigma_x} \text{exp}\left(-\frac{x^2}{2 \sigma_x^2} \right) \mathit{dx} \right] \cdot{} \left[ \int_{- \frac{H}{2}}^\frac{H}{2} \frac{1}{\sqrt{2 \pi} \sigma_y} \text{exp}\left(-\frac{y^2}{2 \sigma_y^2} \right) \mathit{dy} \right]
 = \text{erf}\left( \frac{W}{2 \sqrt{2} \sigma_x} \right) \text{erf}\left( \frac{H}{2 \sqrt{2} \sigma_y} \right).
\end{equation}
Finally, we obtain the $\mathit{ER}$ as $1-P(D)$:
\begin{equation}
 \label{eqn:ER_simple}
 ER = 1 - \text{erf}\left( \frac{W}{2 \sqrt{2} \sigma_x} \right) \text{erf}\left( \frac{H}{2 \sqrt{2} \sigma_y} \right).
\end{equation}
What we do is to simply compute the success rates for the x- and y-axes independently, multiply them, and subtract the product from one.
However, we used several simplifications and assumptions, such as that $\rho$ is negligible, and the appropriate formulations to predict the interaction terms for $\sigma_x$ and $\sigma_y$ are also unknown.
We should thus empirically determine the formulations and evaluate our model's prediction accuracy.

\section{Experiment 1: Cyclic Pointing with Four Movement Angles}
Because recruiting numerous participants helps researchers obtain the central tendency of $\mathit{ER}$s \cite{Yamanaka21hcomp}, we conducted a cyclic pointing experiment on the {\sl Yahoo! Crowdsourcing} platform (\url{https://crowdsourcing.yahoo.co.jp}).
The experimental system was developed with the \texttt{Hot Soup Processor} programming language.
The crowdworkers were asked to download and run an executable file to perform the task.
Our affiliation's IRB-equivalent research ethics team and the crowdsourcing platform approved this study.

\subsection{Task, Design, and Procedure}
\begin{figure}[t]
\centering
\includegraphics[width=0.7\columnwidth]{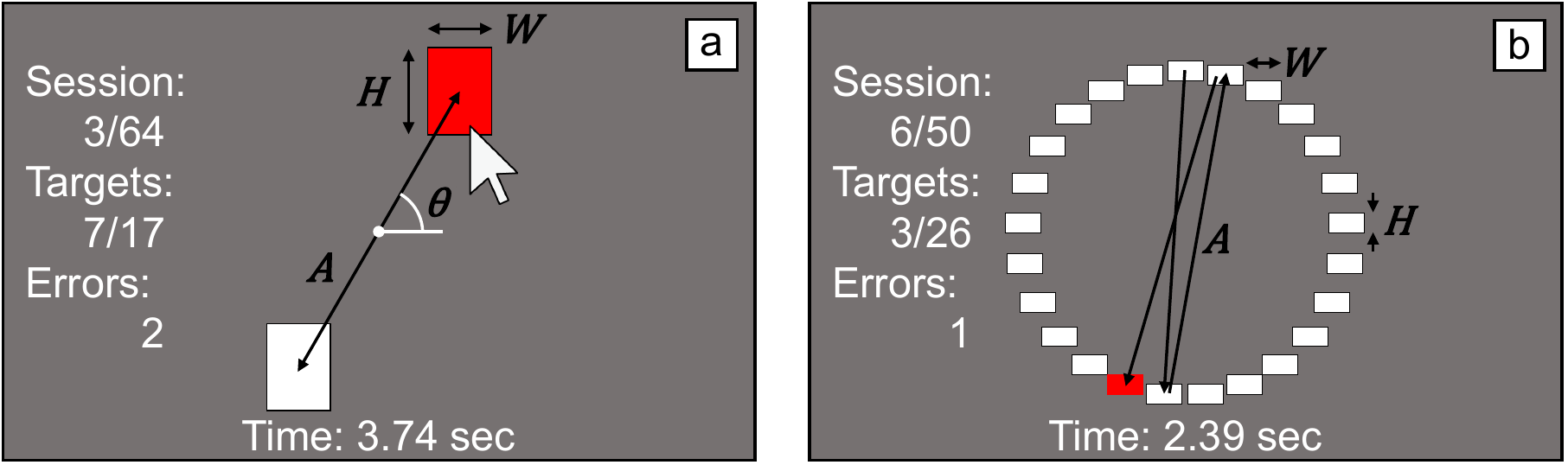}
\caption{Visual stimuli used in Experiments (a) 1 and (b) 2.}
\label{fig:studyScreen}
\end{figure}

In the task window ($1200 \times 1000$ pixels), two rectangular targets were displayed (Figure~\ref{fig:studyScreen}a).
If the workers clicked the target, the red target and white non-target switched colors, and they repeatedly performed this action back and forth.
If the workers missed the target, it flashed yellow and they had to keep trying until successfully clicking it.
The distance between the target centers $A$ was fixed to $550$ pixels.
The movement angle was defined as $\theta$.
A \textit{session} consisted of 17 clicks for a fixed $(W, H, \theta)$ condition.
The first target was on the left side (or the bottom when $\theta = 90^\circ$).

This study was a repeated-measures design with two independent variables: $W$ and $H$ (30, 50, 80, and 120 pixels for both).
While we did not include $\theta$ as an independent variable, we tested four $\theta$s ($0^\circ$, $30^\circ$, $60^\circ$, and $90^\circ$), which included two directions of outbound/return (e.g., rightward and leftward movements, respectively, for when $\theta = 0^\circ$).
We chose these four values so that $\theta$ ranged from horizontal to vertical angles.
While we would like to include more angles, we limited the range to $0^\circ \le \theta \le 90^\circ$ to avoid an overly large number of task conditions.

The order of the 64 $W \times H \times \theta$ conditions was randomized.
The first three out of 17 clicks for each session were eliminated.
Thus, $14_\mathrm{clicks} \times 64_\mathrm{conditions} = 896_\mathrm{trials}$ were recorded for each worker.
Trials in which we observed one or more clicks outside the target were flagged as an error.
The dependent variables were $\sigma_x$, $\sigma_y$, and $\mathit{ER}$.

\subsection{Participants and Recruitment}
We requested no specific PC skills of the crowdworkers, but used the ``White List'' option in the crowdsourcing platform for screening newly created accounts.
This option enabled us to offer the task only to workers who were considered \textit{reliable} on the basis of their previous task history.

To reduce noise introduced by multiple pointing devices, we asked the workers to use a mouse if they had one.
Nevertheless, we explained that any device was acceptable then removed the non-mouse users from the analysis.
To increase ecological validity, the workers were not asked to change the cursor speed or acceleration function.

Once workers accepted the task, they were asked to read the online instructions, which stated that they should perform the pointing task as rapidly and accurately as possible.
After they finished all sessions and completed the questionnaire, they uploaded the log data file to a server to receive payment.

In total, 210 mouse-users completed the task.
The participants' demographics were as follows.
Age: ranging from 20 to 76 years, $M = 45.0$ and $\mathit{SD} = 10.3$.
Gender: 176 were male and 34 were female.
Handedness: 189 were right-handed and 21 were left-handed.
Windows version: 190 used Win 10, 15 Win 7, 4 Win 8, and 1 Vista.
PC usage history: ranged from 0 (less than 1 year) to 40 years, $M = 22.5$ and $\mathit{SD} = 7.21$.

Each worker received JPY 200 ($\sim$USD 1.77).
The time for the pointing task ranged from 7 min 53 sec to 23 min 5 sec ($M=$ 13 min 1 sec).
Thus, the effective hourly payment was JPY 922 ($\sim$USD 7.95) on average.
Note that this effective payment could change depending on other factors such as the time for reading the instructions.

\section{Results of Experiment 1}
\label{sec:results}
\subsection{Outlier Data Screening}
We removed spatial outliers if the distance of the first click position was shorter than $A/2$ \cite{Banovic13,MacKenzie08} to omit clear accidental operations such as double-clicking the previous target.
Another criterion used in these laboratory-based studies was to remove trials in which the click position was more than $2W$ from the target center, but we did not use this criterion, as there are a variety of target-size definitions in our diagonal-movement task with rectangular targets \cite{MacKenzie92twoD}.

To remove extremely fast or slow operations, we used the inter-quartile range ($\mathit{IQR}$) method \cite{Devore11probability}.
The $\mathit{IQR}$ is defined as the difference between the first and third quartiles of the $\mathit{MT}$ for each session.
Trials in which $\mathit{MT}$ was more than $3 \times \mathit{IQR}$ higher than the third quartile or more than $3 \times \mathit{IQR}$ lower than the first quartile were removed.

For participant-level outliers, we calculated the mean $\mathit{MT}$ across all 64 sessions for each worker.
Using each worker's mean $\mathit{MT}$, we again applied the $\mathit{IQR}$ method for each worker.

Among the 188{,}160 trials ($=896_\mathrm{trials} \times 210_\mathrm{workers}$), we removed 2{,}758 trial-level and two participant-level outliers.
Because the outlier workers also exhibited trial-level outliers, the data from 4{,}513 trials were removed (2.40\%), which was similar to previous studies \cite{Findlater17,yamanaka21penalty}.

\subsection{Analyses of Dependent Variables}
\subsubsection{$\sigma_x$}
While ANOVAs can be robust regardless of the data distribution \cite{Dixon2008models, Mena2017non}, it is better to log-transform the data for detecting statistical significance more appropriately.
The log-transformed data from 7 out of 16 conditions ($4_W \times 4_H$) passed the Shapiro-Wilk normality test ($\alpha=0.05$), or 43.75\%.
This normality test was conducted to examine if the 208 workers' $\sigma_x$ data distributed normally, and the results were independent from whether the endpoints were distributed normally.
We used RM-ANOVAs with the Bonferroni $p$-value adjustment method for pairwise tests.
Throughout this paper, for the $F$ statistic, the degrees of freedom were corrected using the Greenhouse-Geisser method when Mauchly's sphericity assumption was violated ($\alpha = 0.05$).

We found significant main effects of $W$ ($F_{2.415,500.0}=7765.2$, $p<0.001$, $\eta_p^2=0.97$) and $H$ ($F_{2.762,571.7}=24.35$, $p<0.001$, $\eta_p^2=0.11$) on $\sigma_x$.
The interaction of $W$ and $H$ was significant ($F_{8.494,1758}=10.33$, $p<0.001$, $\eta_p^2=0.048$).
For all pairwise comparisons for $W$ and $H$, the differences were significant (with $p < 0.01$ at least).
As either $W$ or $H$ increased, $\sigma_x$ increased.
For the interaction effect, all pairwise comparisons showed $p < 0.05$ at least.

\subsubsection{$\sigma_y$}
For the log-transformed $\sigma_y$ data, 6 out of 16 conditions passed the normality test (37.5\%).
We found significant main effects of $W$ ($F_{2.378,492.2}=29.13$, $p<0.001$, $\eta_p^2=0.12$) and $H$ ($F_{2.150,445.1}=6501.3$, $p<0.001$, $\eta_p^2=0.97$) on $\sigma_y$.
The interaction of $W$ and $H$ was significant ($F_{8.124,1681.7}=44.92$, $p<0.001$, $\eta_p^2=0.18$).
For all pairwise comparisons for $W$ and $H$, the differences were significant (with $p < 0.01$ at least), except for $W = 30$ vs. $50$ pixels ($p=1.0$) and $W = 30$ vs. $80$ pixels ($p < 0.05$).
As the $W$ or $H$ increased, $\sigma_y$ increased.

\subsubsection{Error Rate}
We used non-parametric ANOVAs with the aligned rank transform with multi-factor contrast tests for pairwise comparisons \cite{Elkin21ART,Wobbrock11ART} with Holm's method for $p$-value adjustment.
We found significant main effects of $W$ ($F_{3,621}=37.18$, $p<0.001$, $\eta_p^2=0.15$) and $H$ ($F_{3,621}=20.94$, $p<0.001$, $\eta_p^2=0.092$) on the $\mathit{ER}$.
The interaction of $W$ and $H$ was significant ($F_{9,1863}=3.444$, $p<0.001$, $\eta_p^2=0.016$).
As the $W$ or $H$ increased, the $\mathit{ER}$ decreased.

\subsection{Model Fitting}
For the $208_\mathrm{workers} \times 4_W \times 4_H = 3{,}328$ conditions, 2{,}814 data points of $\sigma_x$ passed the normality test, or 84.6\%.
For $\sigma_y$, 2{,}561 data passed (77.0\%).
We found that 2{,}666 data points passed the bivariate normality test (80.1\%).
These rates were lower than those in previous studies \cite{Bi13b,Yamanaka20issFFF}.

\begin{figure}[t]
\centering
\includegraphics[width=0.7\columnwidth]{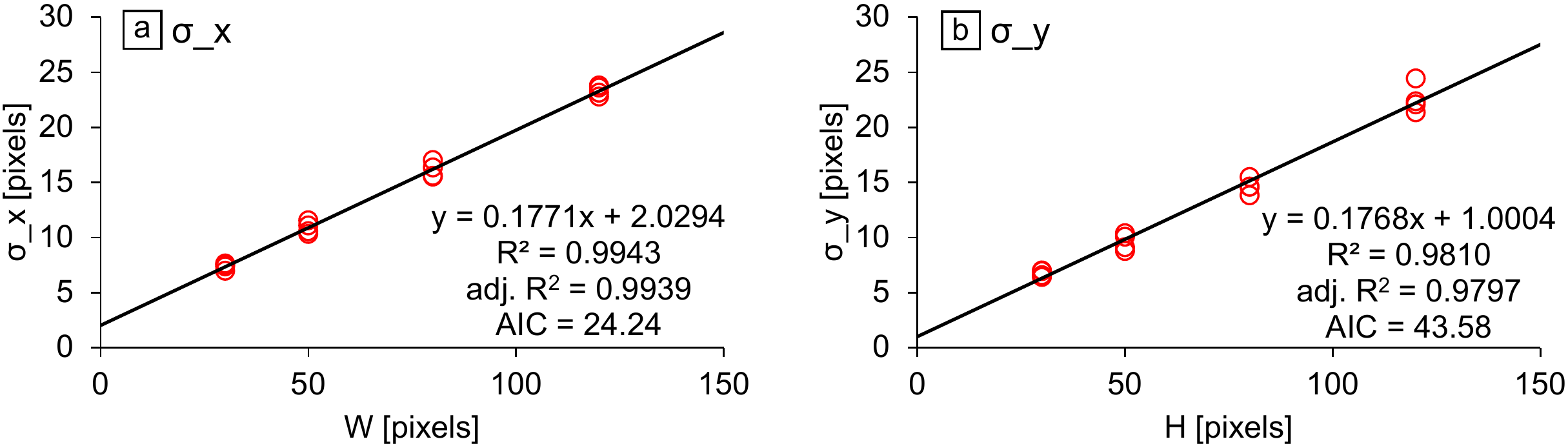}
\caption{Regressions on (a) $\sigma_x$ vs. $W$ and (b) $\sigma_y$ vs. $H$ in Experiment 1.}
\label{fig:ang_sigLinear}
\end{figure}

The $\sigma$s were almost linearly related to the target sizes ($R^2 > 0.98$); see Figure~\ref{fig:ang_sigLinear}.
Thus, using only $W$ for predicting $\sigma_x$ and only $H$ for $\sigma_y$ is possible (Equations~\ref{eqn:sigx_W_proportional} and \ref{eqn:sigy_H_proportional}, respectively).
Yet, as we observed the significant main effects of $W$ and $H$ both for $\sigma_x$ and $\sigma_y$, as well as the interaction of $W$ and $H$, we used the 3-variable formulations (Equation~\ref{eqn:sigx_WH}).

\begin{figure}[t]
\centering
\includegraphics[width=0.7\columnwidth]{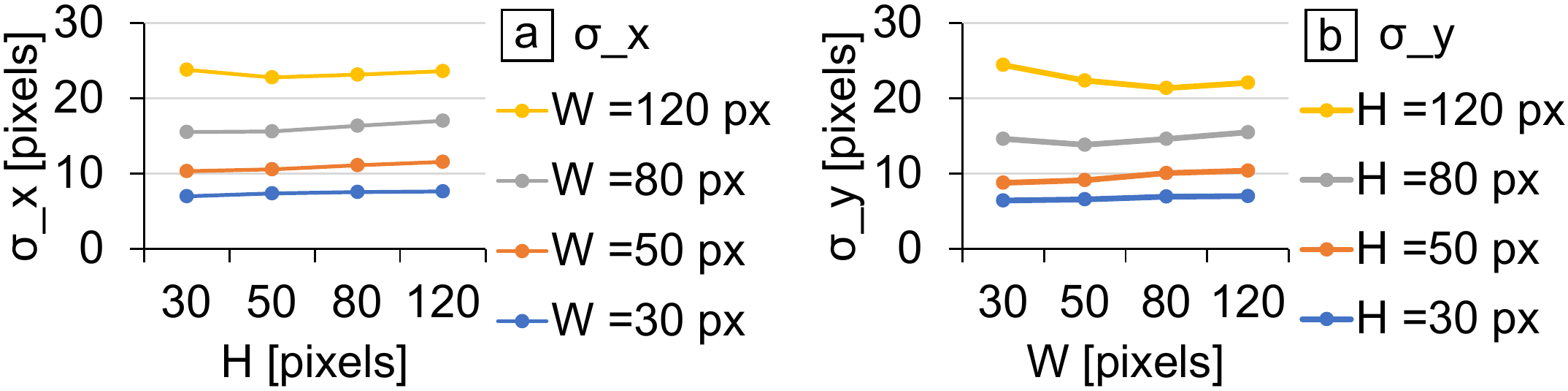}
\caption{Interaction effects of $W$ and $H$ on (a) $\sigma_x$ and (b) $\sigma_y$ in Experiment 1.}
\label{fig:ang_interactionSD}
\end{figure}

As shown in Figure~\ref{fig:ang_interactionSD}a, when $W = 120$ pixels (the yellow line), the largest $\sigma_x$ was observed under the $H = 30$-pixel condition.
In contrast, for the other $W$ conditions, the $\sigma_x$ increased as $H$ increased.
Hence, the positive effect of $W$ on $\sigma_x$ reduced
as $H$ increased; therefore, we added the interaction term of $W$ and $H$ as $(W/H)$.
This interaction was more clearly observed for the $\sigma_y$ result (Figure~\ref{fig:ang_interactionSD}b), and the interaction term on $\sigma_y$ was added as $(H/W)$.
We obtained the fitting results as follows.
\begin{alignat}{4}
\sigma_x &= 0.8407 + 0.1698W + 0.01698H + 0.3949(W/H),& \ \ \ \mathrm{with} \ \ \ R^2 &= 0.9981,& \ \mathrm{adj.}\ R^2 &= 0.9976,& \ \mathit{AIC} &=10.91 \\
\sigma_y &= -1.037 + 0.1508H + 0.02911W + 1.398(H/W),& \ \ \ \mathrm{with} \ \ \ R^2 &= 0.9905,& \ \mathrm{adj.}\ R^2 &= 0.9881,& \ \mathit{AIC} &=36.55
\end{alignat}
For $\sigma_x$, the $\mathit{AIC}$ difference from the 1-variable formulation ($\sigma_x = a+ b\cdot{}W$; see Figure~\ref{fig:ang_sigLinear}a) was greater than 2.
Thus, this 3-variable version was significantly better \cite{Devore11probability} and the two additional terms ($H$ and $W/H$) did not cause overfitting.
This is also true for $\sigma_y$.

\begin{figure}[t]
\centering
\includegraphics[width=0.3\columnwidth]{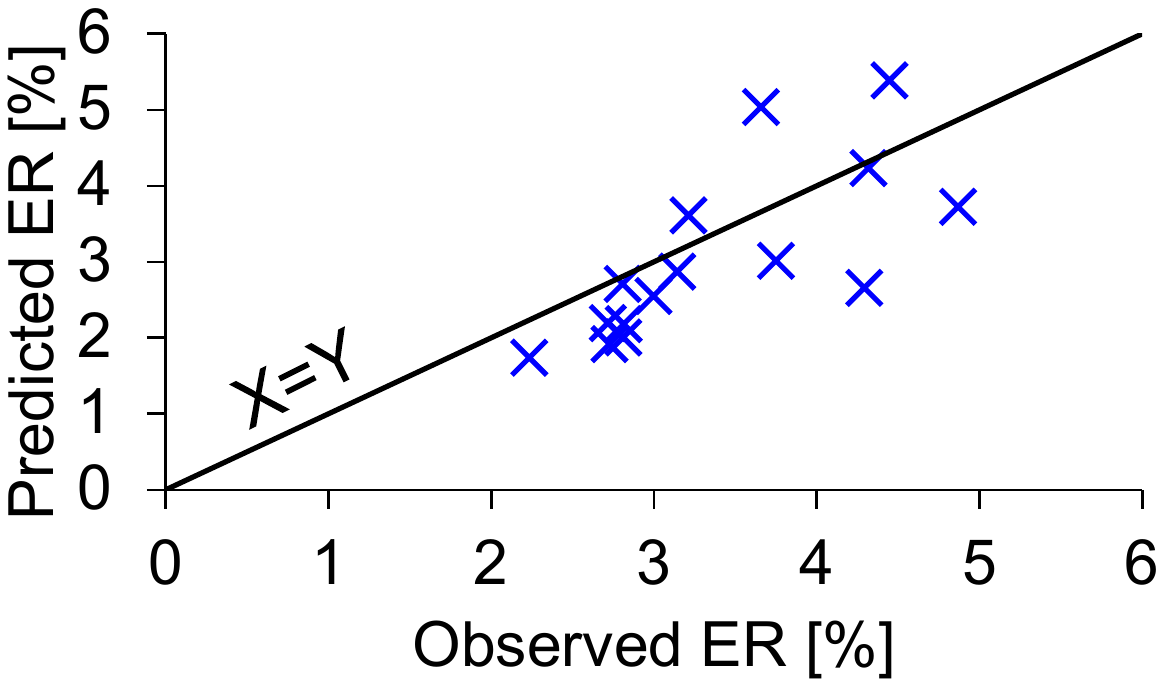}
\caption{Observed vs. predicted $\mathit{ER}$s for the 3-variable formulation in Experiment 1.}
\label{fig:ang_ERcomparison}
\end{figure}

Using these regression expressions, we obtain the predicted $\sigma_x$ and $\sigma_y$ for each target condition then compute the $\mathit{ER}$s by using Equation~\ref{eqn:ER_simple}.
The observed vs. predicted $\mathit{ER}$s are shown in Figure~\ref{fig:ang_ERcomparison}a.
The results of three metrics to evaluate prediction accuracy were $R^2= 0.5315$, $\mathit{MAE} = 0.8270\%$, and $\mathit{RMSE} = 1.059\%$.

These results are, however, worse than those for the $\mathit{ER}$s using 1-variable formulations to predict $\sigma_x$ and $\sigma_y$.
The comparison is shown in Table~\ref{tab:ang_cross} (see the ``All data'' column).
The results are $R^2 = 0.8862$, $\mathit{MAE} = 0.5758\%$, and $\mathit{RMSE} = 0.6259\%$.

\renewcommand{\arraystretch}{1.0}
\begin{table}[t]
\caption{$\mathit{ER}$ prediction accuracy in Experiment 1 to compare 1- vs. 3-variable formulations.}
\label{tab:ang_cross}
\scalebox{0.95}{
\begin{tabular}{cc|c|c|c|c}
\hline
$\sigma$ formulation & Metric & All data & 80\% train : 20\% test & 70\% train : 30\% test & 60\% train : 40\% test \\ \hline 
& $R^2$ & 0.8862 & 0.8482 & 0.8326 & 0.8145 \\ 
\rowcolor[gray]{0.9} \cellcolor[rgb]{1, 1, 1}{1-variable} & $\mathit{MAE}$ [\%] & 0.5758 & 0.6208 & 0.6177 & 0.6596 \\ 
& $\mathit{RMSE}$ [\%] & 0.6259 & 0.6766 & 0.6878 & 0.7474 \\ \hline 
& $R^2$ & 0.5315 & 0.7398 & 0.6648 & 0.5409 \\ 
\rowcolor[gray]{0.9} \cellcolor[rgb]{1, 1, 1}{3-variable} & $\mathit{MAE}$ [\%] & 0.8270 & 0.8712 & 0.8921 & 0.9318 \\ 
& $\mathit{RMSE}$ [\%] & 1.0589 & 1.0513 & 1.1011 & 1.1981 \\ \hline 
\end{tabular}
}
\bigskip\centering
\end{table}
\renewcommand{\arraystretch}{1.0}

\subsection{Cross Validation}
In the above-mentioned results, we used all data from 16 target conditions to be fitted.
To investigate the prediction accuracy for future (unknown) task conditions, we ran shuffle-split cross-validations to evaluate the prediction accuracy when the training data size becomes smaller.

We used three ratios of (train:test) data: (80\%:20\%), (70\%:30\%), and (60\%:40\%).
For example, when the training data size is 70\%, we randomly choose \texttt{ceil}$(0.7 \times 16) = 12$ conditions for computing the coefficients of models then predict the $\mathit{ER}$s for the remaining four conditions.
By comparing the predicted vs. observed $\mathit{ER}$s for the four test data, we obtain $R^2$, $\mathit{MAE}$, and $\mathit{RMSE}$.
We repeatedly carried out this process over 100 iterations to handle the sampling randomness when splitting
the training and test datasets and obtained the average values of the three metrics.
The results are shown in Table~\ref{tab:ang_cross}.

\subsection{Discussion of Experiment 1}
Better fits were observed for the 1-variable formulation regardless of the analyses of all-data or cross-validation.
However, as a limitation of Experiment 1, the number of movement angles were still limited.
This caused an issue that the x- and y-axes distributions had correlations ($\rho$) ranging from $-0.2094$ to $0.01614$ for the 16 target conditions, and the mean was $-0.1078$.
A negative $\rho$ indicates that the spread of endpoints had an ellipse diagonal along the top-right to bottom-left direction on the screen.
This is obviously because we tested only such angles ($0^\circ \le \theta \le 90^\circ$) and is consistent with Ma et al.'s report on rectangular-target pointing with known movement angles \cite{Ma21uist}.

This $\rho$ range is statistically considered ``negligible correlations'' \cite{Hinkle90}, which supported our simplification.
Yet, to check whether the lower fitness was observed due to the lack of $\rho$, we use the correlations for the 16 fitting points in the more complete version of probability computation (Equation~\ref{eqn:SR_original_corr}).
The $R^2$ was $0.5320$ for the all-data analysis, which has only a 0.0005-point difference from the correlation-ignored 3-variable formulation version.
Therefore, integrating the correlations did not improve prediction accuracy.

Furthermore, the target sizes were somewhat large (30 to 120 pixels), which enabled the workers to point to targets with a certain accuracy (less than 5\%; see Figure~\ref{fig:ang_ERcomparison}).
This made it difficult for us to determine if the models could accurately predict $\mathit{ER}$s when it would be higher, e.g., 8\%.
To address these limitations, we conducted Experiment 2.

\section{Experiment 2: Multi-directional Pointing}
To further validate our models, we conducted Experiment 2 with much more variety of angles and smaller target sizes.
We used the same crowdsourcing platform, and the points of difference are described in the following subsections.

\subsection{Participants}
In total, 267 mouse-users completed the task.
Age: ranging from 17 to 73 years, $M = 44.6$ and $\mathit{SD} = 10.1$.
Gender: 229 were male, 36 were female, 2 did not answer.
Handedness: 247 were right-handed and 20 were left-handed.
Windows version: 229 used Win 10, 22 Win 7, 9 Win 11, 6 Win 8, and 1 Vista.
PC usage history: ranged from 1 to 45 years, $M = 23.8$ and $\mathit{SD} = 7.73$.
Each worker received JPY 250 ($\sim$USD 2.26).
The time for the task ranged from 14 min 2 sec to 40 min 8 sec ($M=$ 19 min 45 sec).
The effective hourly payment was JPY 759 ($\sim$USD 6.55) on average.

\subsection{Task, Design, and Procedure}
This task was a modified version of the ISO 9241-9 standard \cite{iso2000} with rectangular targets.
Twenty-five targets appeared in a ring-shaped arrangement, and the first target was located at the top of the window (Figure~\ref{fig:studyScreen}b).
If the participants clicked it, the next target (bottom-left one) turned red, and they successively selected the targets in a pre-programmed order.
The $A$ was fixed to 500 pixels.
A \textit{session} consisted of 26 selections (the top target acted as the start and end ones) for a fixed $(W, H)$ condition.

This study was a repeated-measures design with two independent variables of $W$ and $H$: 12, 18, 26, 36, 48, 62, and 78 pixels for both.
The order of the 49 $W \times H$ conditions was randomized.
The first click for each session was eliminated, and the remaining 25 clicks were analyzed.
Thus, $25_\mathrm{clicks} \times 49_\mathrm{conditions} = 1125_\mathrm{trials}$ were recorded for each participant.
Before the first session, the participants performed a practice session with $A = 400$, $W = 45$, and $H = 31$ pixels, i.e., the condition that was not used in the main 49 sessions.

\section{Results of Experiment 2}
We used the same outlier-detection criteria as in Experiment 1.
Among the 327{,}075 trials ($=1225_\mathrm{trials} \times 267_\mathrm{workers}$), we removed 3{,}465 trial-level and one participant-level outliers.
Because the outlier worker also exhibited trial-level outliers, the data from 4{,}663 trials were removed (1.43\%).

\subsection{Analyses of Dependent Variables}
\subsubsection{$\sigma_x$}
We found that the log-transformed $\sigma_x$ data passed the normality test for 18 out of 49 conditions ($7_W \times 7_H$), or 36.7\%.
An RM-ANOVA showed significant main effects of $W$ ($F_{2.987,791.6}=9716$, $p<0.001$, $\eta_p^2=0.97$) and $H$ ($F_{4.707,1273}=6.879$, $p<0.001$, $\eta_p^2=0.025$) on $\sigma_x$.
The interaction of $W$ and $H$ was significant ($F_{29.78,7891}=13.08$, $p<0.001$, $\eta_p^2=0.047$).
For all pairwise comparisons for $W$ and $H$, the differences were significant (with $p < 0.05$ at least).
For the interaction effect, all pairwise comparisons showed $p < 0.05$.

\subsubsection{$\sigma_y$}
We found that the log-transformed $\sigma_y$ data passed the normality test for 14 conditions, or 28.6\%.
An RM-ANOVA showed significant main effects of $W$ ($F_{4.412,1169}=48.14$, $p<0.001$, $\eta_p^2=0.18$) and $H$ ($F_{2.679,710.0}=6317$, $p<0.001$, $\eta_p^2=0.96$) on $\sigma_y$.
The interaction of $W$ and $H$ was significant ($F_{29.56,7833}=26.60$, $p<0.001$, $\eta_p^2=0.091$).
For all pairwise comparisons for $W$ and $H$, the differences were significant (with $p < 0.05$ at least).
For the interaction effect, all pairwise comparisons showed $p < 0.05$.

\subsubsection{Error Rate}
A non-parametric ANOVA with aligned rank transform showed significant main effects of $W$ ($F_{6,1590}=23.10$, $p<0.001$, $\eta_p^2=0.080$) and $H$ ($F_{6,1590}=137.1$, $p<0.001$, $\eta_p^2=0.34$) on $\mathit{ER}$.
The interaction of $W$ and $H$ was significant ($F_{36,9540}=4.909$, $p<0.001$, $\eta_p^2=0.018$).
The differences were not significant for $W = $ (48, 62), (48, 78), and (62, 78) pixels, while the other $W$ pairs showed significant differences ($p < 0.05$).
This was also true for the same $H$s.
Thus, the significant $\mathit{ER}$ differences disappeared as $W$ and $H$ increased.

When either $W$ or $H$ was small, it had a stronger effect to increase the $\mathit{ER}$.
For example, when $W = 12$ pixels, the $\mathit{ER}$ for $H = 12$ showed a significant pairwise difference only from $H = 78$ pixels (7.612 vs. 6.073\%, respectively; $p < 0.05$).
In comparison, when $W = 78$ pixels, the $\mathit{ER}$ for $H = 12$ showed $p<0.001$ for all the other $H$s, ranging from 0.835\% ($H = 12$) to 9.079\% ($H = 78$).
This clearly showed the interaction effect of $W$ and $H$ on $\mathit{ER}$.

\subsection{Model Fitting}
For the $266_\mathrm{workers} \times 7_W \times 7_H = 13{,}034$ conditions, 11{,}764 data points of $\sigma_x$ passed the normality test, or 90.3\%.
For $\sigma_y$, 11{,}690 data passed (89.7\%).
We found that 11{,}955 data points passed the bivariate normality test (91.7\%).
The assumption that the endpoints are distributed normally was more supported than in Experiment 1.

\begin{figure}[t]
\centering
\includegraphics[width=0.7\columnwidth]{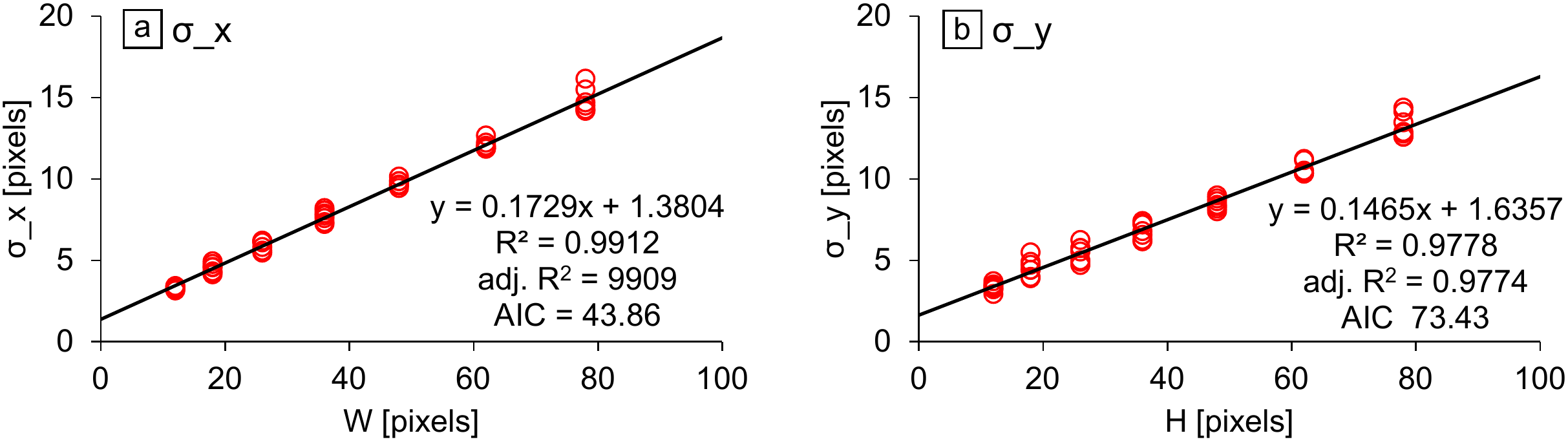}
\caption{Regressions on (a) $\sigma_x$ vs. $W$ and (b) $\sigma_y$ vs. $H$ in Experiment 2.}
\label{fig:mul_sigLinear}
\end{figure}

\begin{figure}[t]
\centering
\includegraphics[width=0.7\columnwidth]{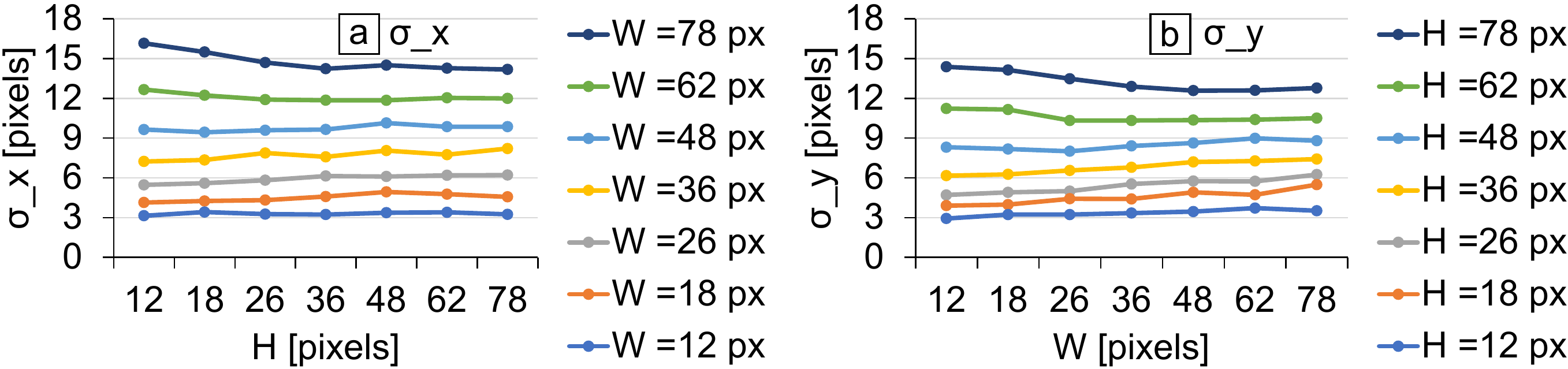}
\caption{Interaction effects of $W$ and $H$ on (a) $\sigma_x$ and (b) $\sigma_y$ in Experiment 2.}
\label{fig:mul_interactionSD}
\end{figure}

The endpoint variabilities were almost linearly related to the target sizes ($R^2 > 0.97$); see Figure~\ref{fig:mul_sigLinear}.
As we observed a similar interaction effect between $W$ and $H$ as in Experiment 1 (see Figure~\ref{fig:mul_interactionSD}), we examined the 3-variable formulations with the interaction terms of $(W/H)$ and $(H/W)$.
The following coefficients were obtained.
\begin{eqnarray}
\sigma_x = 0.8713 + 0.1614W + 0.01273H + 0.3162(W/H),\ \ \ \mathrm{with} \ \ \ R^2 = 0.9944,\ \mathrm{adj.}\ R^2 = 0.9941,\ \mathit{AIC}=25.25 \\
\sigma_y = 0.7221 + 0.1309H + 0.02284W + 0.4282(H/W),\ \ \ \mathrm{with} \ \ \ R^2 = 0.9882,\ \mathrm{adj.}\ R^2 = 0.9874,\ \mathit{AIC}=53.98
\end{eqnarray}

\begin{figure}[t]
\centering
\includegraphics[width=0.3\columnwidth]{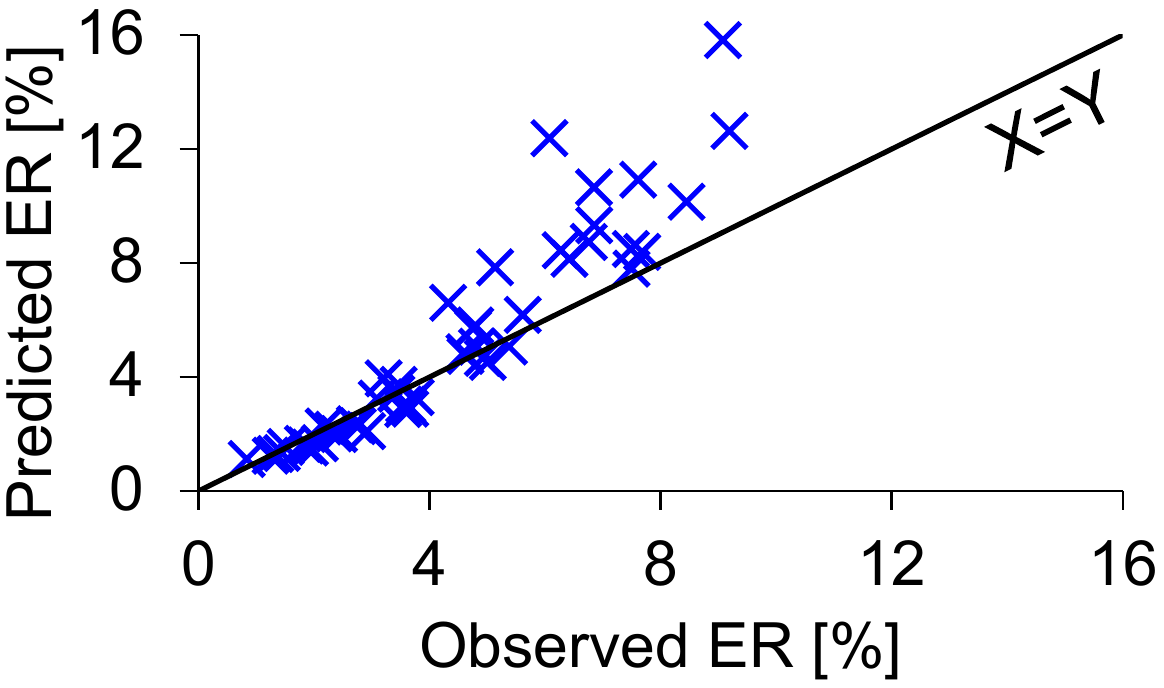}
\caption{Observed vs. predicted $\mathit{ER}$s for the 3-variable formulation in Experiment 2.}
\label{fig:mul_ERcomparison}
\end{figure}

The observed vs. predicted $\mathit{ER}$s are shown in Figure~\ref{fig:mul_ERcomparison}.
The results of the prediction-accuracy metrics are summarized in Table~\ref{tab:mul_cross}.

\subsection{Cross Validation}
The $\mathit{ER}$s ranged from 0.835 to 9.19\% (Figure~\ref{fig:mul_ERcomparison}), which were more widely spread than in Experiment 1 (ranging from 2.23 to 4.87\%).
This enables us to evaluate the models' validity more thoroughly.
As shown in Table~\ref{tab:mul_cross}, the 3-variable formulations exhibited better prediction accuracy in terms of $R^2$ regardless of the training data size.

\renewcommand{\arraystretch}{1.0}
\begin{table}[t]
\caption{$\mathit{ER}$ prediction accuracy in Experiment 2 to compare 1- vs. 3-variable formulations.}
\label{tab:mul_cross}
\scalebox{0.95}{
\begin{tabular}{cc|c|c|c|c}
\hline
$\sigma$ formulation & Metric & All data & 80\% train : 20\% test & 70\% train : 30\% test & 60\% train : 40\% test \\ \hline
& $R^2$ & 0.8189 & 0.8310 & 0.8229 & 0.8182 \\ 
\rowcolor[gray]{0.9} \cellcolor[rgb]{1, 1, 1}{1-variable} & $\mathit{MAE}$ [\%] & 1.0853 & 1.1781 & 1.1789 & 1.2568 \\ 
& $\mathit{RMSE}$ [\%] & 1.8260 & 1.8575 & 1.8645 & 2.0179 \\ \hline 
& $R^2$ & 0.8810 & 0.8943 & 0.8730 & 0.8627 \\ 
\rowcolor[gray]{0.9} \cellcolor[rgb]{1, 1, 1}{3-variable} & $\mathit{MAE}$ [\%] & 1.0919 & 1.0597 & 1.1883 & 1.2283 \\ 
& $\mathit{RMSE}$ [\%] & 1.8327 & 1.7008 & 1.9248 & 2.0459 \\ \hline 
\end{tabular}
}
\bigskip\centering
\end{table}
\renewcommand{\arraystretch}{1.0}

\subsection{Discussion of Experiment 2}
The 3-variable formulations showed better fits than the 1-variable ones in terms of $R^2$.
Because we used much smaller sizes of $W$ and $H$ than in Experiment 1, the perpendicular-axis target size affected the $\sigma$s more strongly, which forced us to use the 3-variable formulations to accurately predict $\mathit{ER}$s.

For the 49 target conditions, the $\rho$ of endpoints ranged from $-0.01488$ to $0.04629$, and the mean was $0.01607$.
The absolute value was considerably smaller than that in Experiment 1 (mean $\rho$ was $-0.1078$).
When we used the $\rho$s for the 49 fitting points in the more complete version of probability computation (Equation~\ref{eqn:SR_original_corr}), the model showed $R^2 = 0.8810$ for the all-data analysis, which is equal to the $\rho$-ignored version (see Table~\ref{tab:mul_cross}), while there was in actuality a $<$0.0001-point difference.
Therefore, when assuming that users aim for the target in various movement angles, ignoring the correlations does not damage prediction accuracy, which supported our simplification.

\section{Conclusions and Future Work}
The appropriate $\sigma$ formulation changed in the two experiments.
In Experiment 1, using the simple 1-variable formulation (Equations~\ref{eqn:sigx_W_proportional} and \ref{eqn:sigy_H_proportional}) showed the best fit in accordance with the cross-validation.
In Experiment 2, however, the 3-variable formulation (Equation~\ref{eqn:sigx_WH}) achieved better fits.
As we used much smaller target sizes in Experiment 2, the perpendicular-axis size affected the $\mathit{ER}$; thus, accurate prediction required considering the endpoint-variability changes depending on both $W$ and $H$.

As we explained in Experiment 1, when the $\mathit{ER}$s are always low, we do not need to use the 3-variable formulations.
Yet, when designers have to use smaller targets due to, e.g., the space limitation of webpages, the necessity to predict $\mathit{ER}$s would increase.
Hence, the 3-variable formulation will work well in more general cases.

The findings and model validation in this study were limited to the extent of our experimental design, such as the target sizes we used.
While we assumed that the target distance does not affect the endpoint variability in accordance with previous studies \cite{Bi13a,Huang18error1D,Yamanaka20issFFF}, this assumption does not hold when users exhibit ballistic movements \cite{Beggs74,Grossman05prob}.
Our instruction to select a target as rapidly and accurately as possible is just one strategy from among various speed-accuracy balances.
Using only mice was another limitation of this study; thus, we plan to evaluate our models with other input methods such as touch-based pointing where other factors would affect endpoint variability \cite{Bi16}.

Our next step is to analyze the prediction accuracy of our models when the frequencies of movement angles are known.
For example, for a target near the top edge of the screen, it is less likely to move the cursor with downward movements; thus, the frequencies of other angles (upward, leftward, diagonally upward, etc.) relatively increase.
Under such a condition, the $\mathit{ER}$s would be more sensitively affected by $W$ rather than $H$, because the frequency of vertical movements are low.
We plan to investigate how we should give weights to the probabilities of each movement angle to accurately predict the overall $\mathit{ER}$.

When we can record the cursor trace on a certain webpage or application, there is another challenge to be addressed, i.e., determine the initial cursor position from continuous movements.
Chapuis et al. described the difficulty in this data processing to segment cursor trajectory \cite{Chapuis07wild}.
They also showed that each pointing did not exhibit a straight-forward motion towards a target; some were linear but others were zig-zag and curved.
Thus, our models, which do not use movement angles, are perhaps useful for practical purposes.

\bibliographystyle{ACM-Reference-Format}
\bibliography{sample-base}

\end{document}